\documentclass[11pt,a4paper,english,nofootinbib,superscriptaddress]{revtex4}
\usepackage{lmodern}

\usepackage[T1]{fontenc}
\usepackage[latin9]{inputenc}
\usepackage{babel}
\usepackage{color}
\usepackage{cancel}
\usepackage{amsmath}
\usepackage{graphicx}
\usepackage{amssymb}
\usepackage{esint}
\usepackage[unicode=true, pdfusetitle,
bookmarks=true,bookmarksnumbered=false,bookmarksopen=false,
breaklinks=false,pdfborder={0 0 1},backref=false,colorlinks=false]
{hyperref}
\usepackage{amsmath}
\usepackage{amssymb}
\usepackage[compat=1.0.0]{tikz-feynman}
\usepackage{float}

\def\b{\begin{equation}} \def\e{\end{equation}}
\def\bd{\begin{displaystyle}} \def\ed{\end{displaystyle}}
\def\ba{\begin{array}} \def\ea{\end{array}}

\def\bee{\begin{enumerate}}
    \def\eee{\end{enumerate}}

\def\1{\mbox{I\hspace{-.15em}1}}

\def\b{\begin{equation}}
\def\e{\end{equation}}
\def\bee{\begin{enumerate}}
    \def\eee{\end{enumerate}}

\makeatletter
\usepackage{latexsym}\usepackage{bm}

\makeatother

\begin{document}

    \title{Dilaton black holes with power law electrodynamics}
    \author{M. Dehghani} \email{e-mail: m.dehghani@razi.ac.ir }
    \affiliation{\it Department of Physics, Razi University,
        Kermanshah, Iran.}

    \author {M.R. Setare}
    \email{e-mail: rezakord@ipm.ir }
     \affiliation{\it Department of Physics, Razi University,
        Kermanshah, Iran.}
    \begin{abstract}
        In this article, the new black hole solutions to the
        Einstein-power-Maxwell-dilaton gravity theory have been investigated
        in a four-dimensional space-time. The coupled scalar,
        electromagnetic and gravitational field equations have been solved
        in a static and spherically symmetric geometry. It has been shown
        that dilatonic potential, as the solution to the scalar field
        equation, can be written in the form of a generalized Liouville
        potential. Also, three classes of novel charged dilaton black hole
        solutions, in the presence of power law nonlinear electrodynamics,
        have been constructed out which are asymptotically non-flat and
        non-AdS. The conserved and thermodynamic quantities have been
        calculated from geometrical and thermodynamical approaches,
        separately. Since the results of these two alternative approaches
        are identical one can argue that the first law of black hole
        thermodynamics is valid for all of the new black hole solutions. The
        thermodynamic stability or phase transition of the black holes have
        been studied, making use of the canonical ensemble method. The
        points of type-1 and type-2 phase transitions as well as the ranges
        at which the black holes are stable have been indicated by
        considering the heat capacity of the new black hole solutions. The
        global stability of the black holes have been studied through the
        grand canonical ensemble method. Regarding the Gibbs free energy of
        the black holes, the points of Hawking-Page phase transition and
        ranges of the horizon radii at which the black holes are globally
        stable have been determined.\\


    \end{abstract}
    \maketitle
    \setcounter{equation}{0}
    \section{\textbf{Introduction}}

    It seems that, at least at the high-energy regime, Einstein's action
    alone is not sufficient to give to describe the Universe completely.
    This action has been modified by the superstring terms which are
    scalar tensor in nature. The low-energy limit of the string theory
    leads to the Einstein gravity coupled to a scalar dilaton field
    \cite{ewi, gib}. Since ever the new string black hole solutions have
    been found, there has been strong interest in studying the exact
    solutions to the scalar coupled general relativity known as the
    Einstein-dilaton gravity theory. In the presence of dilaton field
    the asymptotic behavior of the solutions is changed to be neither
    flat nor (A)dS \cite{2017, 20172}. Although, existence of the
    dilatonic black holes in the space-times with negative cosmological
    constant circumvents the no-hair conjecture, which originally stated
    that a black hole should be characterized only by its mass, angular
    momentum and electric charge \cite{ru, ru2}, it has been shown by
    many authors that Einstein's gravity theory with a coupled scalar
    field admits exact hairy black hole solutions in three-, four- and
    higher-dimensional space-times \cite{dilaton}-\cite{za2}.

    The studies on the black holes, as the thermodynamic systems, date
    back to the most outstanding achievements of Hawking and Bekenstein
    in the context of geometrical physics. According to the laws of
    black hole thermodynamics the black hole temperature and entropy are
    related to the geometrical quantities such as black hole horizon
    area and surface gravity, respectively. Also, the black hole entropy
    and temperature together with the black hole mass (energy) satisfy
    the first law of black hole thermodynamics \cite{1}-\cite{haw}.

    Besides, thermodynamic stability or phase transition of the black
    holes is the other important issue that has attracted a lot of
    attentions in recent years.  There are alternative approaches for
    analyzing the thermal stability or thermodynamic phase transition of
    the black holes. Geometrical thermodynamics is an interesting way
    for investigation of black hole phase transition. In this method,
    divergence points of thermodynamical Ricci scalar provide some
    information related to thermodynamic phase transition points (see
    \cite{hegt} and references therein). Making use of the grand
    canonical ensemble method and noting the signature of the
    determinant of the Hessian metrics one can find some information
    about the thermodynamic stability of the physical black holes
    \cite{ref2, ref1, ref3}. Also, the canonical ensemble method is an
    interesting way for studying the black hole stability which is based
    on the behavior of the black hole heat capacity. It is argued that
    roots and divergence points of the black hole heat capacity are
    representing the two types of phase transitions. In addition, the
    signature of black hole heat capacity with the black hole charge as
    a constant, enables one to study the thermal stability of the black
    holes \cite{he22, de1, de2}. In this paper, we want to study thermal
    stability and phase transition in the context of canonical ensemble
    method as the more fundamental theory.

    Recent studies on the thermodynamics of black holes have shown that
    there is a correspondence between the gravitating fields in anti-de
    Sitter (AdS)  space-time and conformal field theory (CFT) living on
    the boundary of the AdS space-time. Thus the thermodynamic
    properties of black holes in AdS spaces can be identified based on
    the AdS/CFT correspondence in the high temperature limit \cite{wi,
        wi2, wi3}.

    The appearance of infinite electric field and self-energy for the
    pointlike charged particles, as the famous challenge of Maxwell's
    theory of classical electrodynamics, was the initial motivation for
    introducing the various models of nonlinear electrodynamics. The
    first model of nonlinear electrodynamics, which restricts the
    electric field of a point charge to an upper bound, was proposed by
    Born and Infeld in 1935 \cite{bi1, bi2}. It was shown that
    Born-Infeld nonlinear electrodynamics is not the only modification
    of the linear Maxwell field which keeps the electric field of a
    charged point particle finite at the origin, and other types of
    nonlinear Lagrangian such as exponential and logarithmic nonlinear
    electrodynamics can play the same role \cite{so, kazemi, hajkh,
        log}. Nowadays, the theory of nonlinear electrodynamics has got a
    lot of attentions, and has provided an interesting research area in
    the context of geometrical physics and specially in the studying of
    static and rotational black holes. Utilizing the Born-Infeld and the
    other types of nonlinear electrodynamics such as logarithmic,
    exponential and power-law electrodynamics, lead essentially to some
    new black hole solutions with the physical and thermodynamical
    properties affected by the model of electrodynamics under
    consideration \cite{de1, de2} and \cite{kazemi}-\cite{prd}. In the
    four-dimensional space-times, the Lagrangian of Maxwell's theory
    remains invariant under conformal transformations in the form of
    $g_{\alpha\beta}\rightarrow \Omega^2 g_{\alpha\beta}$. Breaking down
    of the conformal symmetry in the space-times with the dimensions
    other than four is the other challenge of Maxwell's theory of
    electrodynamics. Power-law theory of nonlinear electrodynamics is
    the only model of electrodynamics which preserves its conformally
    invariant property in the space-times with arbitrary dimensions. It
    has been shown that the Lagrangian of power-law nonlinear
    electrodynamics is invariant under the conformal transformations in
    the three- four- and higher-dimensional space-times provided that
    the power is chosen equal to one-fourth of the space-time dimensions
    \cite{ref2, ash, ko3}. Since the action of Einstein-dilaton gravity
    theory is related to that of scalar-tensor gravity theory via
    conformal transformations, this conformal symmetry of the power-law
    nonlinear electrodynamics makes it more interesting to be considered
    in the context of geometrical physics and specially in the framework
    of Einstein-dilaton gravity theory \cite{jor1}-\cite{jor4}.

    The main goal of this work is to obtain the novel exact black hole
    solutions to the Einstein-power-Maxwell-dilaton gravity theory and
    to investigate the physical and thermodynamical properties of the
    solutions. Also, to check the validity of the thermodynamical first
    law as well as to perform a thermal stability or phase transition
    analysis for the new black hole solutions. Indeed, this work can be
    regarded as the extension of my previous one presented in ref.
    \cite{IJMPD}, named as the Einstein-Maxwell-dilaton gravity theory,
    to the case of nonlinear electrodynamics by considering the power
    law Maxwell field.

    The paper is structured based on the following order. In Sec. II, by
    starting from a suitable four-dimensional Einstein dilatonic action
    coupled to the power law nonlinear electrodynamics, we obtained the
    related field equations. We have solved the equations of the scalar,
    electromagnetic and tensor fields in a static spherically symmetric
    geometry and showed that the dilatonic potential can be written as
    the linear combination of Liouville-type potentials. Also, two
    classes of new black hole solutions, as the exact solutions to the
    Einstein-power-Maxwell-dilaton gravity theory have been constructed
    out, which are asymptotically non-flat and non-AdS. Sec. III is
    devoted to study of the thermodynamic properties of the novel
    charged black hole solutions. The black hole total charge and mass,
    as the conserved quantities, as well as the entropy and temperature
    associated with the black hole horizon have been obtained. Also, the
    electric potential of the black holes, relative to a reference point
    located at infinity relative to the horizon, has been obtained. In
    addition, through a Smarr-type mass formula, we have obtained the
    black hole mass as a function of the extensive parameters, charge
    and entropy. The intensive parameters, temperature and electric
    potential, conjugate to the extensive parameters, have been
    calculated from thermodynamical methods. Compatibility of the
    results of geometrical and thermodynamical approaches confirms the
    validity of the first law of black hole thermodynamics, for all
    classes of the new black hole solutions. Sec. IV is dedicated to
    investigation of the local stability or phase transition of the
    black holes. Making use of the canonical ensemble method and
    regarding the black hole heat capacity, with the black hole charge
    as a constant, a black hole stability analysis has been performed
    and the points of type-1 and type-2 phase transitions as well as the
    ranges at which the black holes are locally stable have been
    determined, precisely. A black hole global stability analysis has
    been presented in Sec. V. Through calculation of the black hole
    Gibbs free energy the points of the Hawking-Page phase transition
    and the ranges at which our black holes are globally stable have
    been characterized. Some concluding results are summarized and
    discussed in Sec. VI.

    \setcounter{equation}{0}
    \section{\textbf{The field equations and the black hole solutions}}

    We start with the action of the four-dimensional charged black holes
    in the Einstein gravity theory coupled to a dilatonic potential. It
    can be written in the following general form \cite{IJMPD, he} \b\label{2.1}
    I=\frac{1}{16\pi}\int
    \sqrt{-g}d^4x\left[{\cal{R}}-V(\phi)-2g^{\mu\nu}\nabla_\mu
    \phi\nabla_\nu \phi+{\cal{L}}({\cal{F}}, \phi)\right].\e Here,
    ${\cal{R}}$ is the Ricci scalar. $\phi$ is the scalar field coupled
    to itself via the functional form $V(\phi)$. The last term is the
    coupled scalar-electrodynamic lagrangian. Making use of the power
    law nonlinear electrodynamics and in terms of the
    scalar-electromagnetic coupling constant $\alpha$, It can be written
    in the following form \cite{de1, de2, kz1} \b {\cal{L}}({\cal{F}},
    \phi)= \left(-{\cal{F}}e^{-2\alpha \phi}\right)^p,\e where,
    ${\cal{F}}=F^{\mu\nu}F_{\mu\nu}$ being the Maxwell invariant. In
    terms of the electromagnetic potential, $A_\mu$, $F_{\mu\nu}$ is
    defined as $F_{\mu\nu}=\partial_\mu A_\nu-\partial_\nu A_\mu$ and
    power $p$ is known as the nonlinearity parameter. It is expected
    that in the case $p=1$ the results of this theory reduce to the
    Einstein-Maxwell-dilaton gravity theory. Now, by varying the action
    \eqref{2.1}, we get the following field equations

    \b2{\cal{R}}_{\mu\nu}=V(\phi)g_{\mu\nu}+4\nabla_\mu \phi\nabla_\nu
    \phi+\left[(2p-1)g_{\mu\nu}+\frac{2p}{{\cal{F}}}F_{\mu\alpha}F_{\nu}^{\;\alpha}\right]{\cal{L}}({\cal{F}},
    \phi),\e

    \b\nabla_\mu\left[{\cal{L}}_{{\cal{F}}}({\cal{F}},
    \phi)F^{\mu\nu}\right]=0,\;\;\;\;\;\;\;\;{\cal{L}}_{{\cal{F}}}({\cal{F}},
    \phi)\equiv\frac{\partial}{\partial{\cal{F}}}{\cal{L}}({\cal{F}},
    \phi),\e

    \b 4\Box \phi=\frac{d V(\phi)}{d\phi}+2\alpha p {\cal{L}}({\cal{F}},
    \phi),\;\;\;\;\;\;\;\; \phi=\phi(r).\e for the gravitational,
    electromagnetic and scalar field equations, respectively.

    We consider the following ansatz as the four-dimensional spherically
    symmetric solution to the gravitational field equations
    \b ds^2=-W(r)dt^2+\frac{1}{W(r)}dr^2+r^2R^2(r)\left(d\theta^2+\sin^2\theta d\varphi^2
    \right),\e here, $W(r)$ and $R(r)$ are two unknown functions of
    $r$ to be determined.

    Noting the fact that the only nonzero component of the
    electromagnetic field is $F_{tr}$ and assuming as a function of $r$,
    we have $ {\cal{F}}=-2(F_{tr}(r))^2=-2(-\partial_rA_t(r))^2$. In
    overall the paper, prime means derivative with respect to the
    argument. Making use of (II.6) in (II.3), we arrived at the following
    explicit form of the gravitational equations \b
    rR(r)W''(r)+2\left[R(r)+r
    R'(r)\right]W'(r)+rR(r)\left([V(\phi)-{\cal{L}}({\cal{F}},
    \phi)\right)]=0,\e $$ rR(r)W''(r)+2\left[R(r)+r
    R'(r)\right]W'(r)+4\left[rR''(r)+2R'(r)+rR(r)\phi'^2(r)\right]W(r)$$\b+rR(r)\left[V(\phi)-{\cal{L}}({\cal{F}},
    \phi)\right]=0,\e
    $$2rR(r)\left[R(r)-rR'(r)\right]W'(r)+2\left[\left(R(r)\right)^2+r^2R''(r)R(r)+4rR'(r)R(r)+r^2\left(R'(r)\right)^2\right]W(r)$$
    \b\;\;\;\;\;\;\;\;\;\;\;\;\;\;\;\;\;\;\;\;\;\;\;\;\;\;\;\;\;\;\;\;\;\;\;\;\;\;\;\;+r^2R^2(r)\left[V(\phi)+(2p-1){\cal{L}}({\cal{F}},\phi)\right]-2=0,\e
    for $tt$, $rr$ and $\theta\theta\;(\varphi\varphi)$  components,
    respectively. Subtracting Eq.(II.7) from Eq.(II.8) results in \b r
    R''(r)+2R'(r)+rR(r)\phi'^2(r)=0.\e The differential equation (II.10)
    can be written in the following form \b \frac{2}{r}\frac{d}{dr}\ln
    R(r)+\frac{d^2}{dr^2}\ln R(r)+\left(\frac{d}{dr}\ln
    R(r)\right)^2+\phi'^2(r)=0.\e From Eq.(II.11), one can argue that
    $R(r)$ must be an exponential function of $\phi(r)$. Therefore, we
    can write $R(r)=e^{\beta\phi(r)}$ in Eq.(II.11), and show that
    $\phi=\phi(r)$ satisfies the following differential equation \b
    \beta\phi''+(1+\beta^2)\phi'^2+\frac{2\beta}{r}\phi' =0.\e

    The solution of Eq.(II.12), in terms of a positive constant $b$, can
    be written as \b \phi(r)=\gamma \ln\left(
    \frac{b}{r}\right),\;\;\;\;\;
    \mbox{with}\;\;\;\;\;\gamma=\beta(1+\beta^2)^{-1}.\e Here, we are
    interested on studying the effects of the exponential solution (i.e.
    $R(r)=e^{\beta\phi(r)}$) with both $\beta=\alpha$ and
    $\beta\neq\alpha$ cases on the thermodynamics behavior of the
    four-dimensional nonlinearly charged dilatonic black hole solutions.
    The cases of $\beta=\alpha$ and $\beta\neq\alpha$, with Maxwell's
    electromagnetic theory, have been considered in refs. \cite{dilaton,
        IJMPD, 4dRG}. Here, we are interested to extend this idea to the
    charged black hole solutions in the presence of power law nonlinear
    electrodynamics. To do so, we proceed to solve the field equations,
    making use of the scalar fields given by Eq.(II.13).

    Regarding these solutions together with Eq.(II.6), the solution to
    the electromagnetic field equation (II.4) can be written in the
    following form \b \left\{\begin{array}{ll} A_t(r)=
        \frac{q}{B-1}\;r^{1-B},\\\\

        F_{tr}(r)=q\; r^{-B},
    \end{array} \right.\e
    where, $B=\frac{2}{2p-1}\left[1+\gamma(\alpha p-\beta)\right]$ with
    $p\neq\frac{1}{2}$ and $q$ is an integration constant related to the
    total electric charge of the black hole. It will be calculated in
    the following section. A notable point is that in order to the
    potential function $A_t(r)$ be physically reasonable (i.e. zero at
    infinity), the condition $B>1$ must be fulfilled. In the absence of
    dilaton field (i.e $\beta=0=\gamma$) this condition reduces as
    $\frac{2p-3}{2p-1}<0$ or equivalently $\frac{1}{2}<p<\frac{3}{2}$,
    which is just the condition encountered in ref.\cite{de2, stete}.

    Now, Eq.(II.9) can be rewritten as  \b
    W'(r)-\frac{1-2\beta\gamma}{r}W(r)+\frac{r}{2(1-\beta\gamma)}\left[\frac{2}{r^2R^2(r)}-V(\phi)-(2p-1){\cal{L}}({\cal{F}},
    \phi)\right]=0.\e For solving this equation for the metric function
    $W(r)$, we need to calculate the functional form of $V(\phi(r))$ as
    a function of the radial coordinate. For this purpose we return to
    the scalar field equation (II.5). It can be written as \b \frac{d
        V(\phi)}{d\phi}+\frac{4\gamma}{r}\left(W'(r)-\frac{1-2\beta\gamma}{r}W(r)\right)+2\alpha
    p {\cal{L}}({\cal{F}}, \phi)=0.\e Combination of the coupled
    differential equations (II.15) and (II.16) leads to the following
    first order differential equation for the scalar potential \b
    \frac{d V(\phi)}{d\phi}-2\beta V(\phi)+2\left[\alpha
    p-\beta(2p-1)\right]{\cal{L}}({\cal{F}},
    \phi)+\frac{4\beta}{r^2R^2(r)}=0.\e The solution to the differential
    (II.17) can be written as the generalized form of the Liouville
    scalar potential. That is \b V(\phi)=\left\{\begin{array}{ll}
        2(\Lambda+\lambda_1) e^{2\phi}+2\lambda_2\;\phi e^{2\phi}+2\lambda_3\;e^{2\beta_0\phi},\;\;\;\;\; \mbox{for}\;\;\;\beta=1 ,\\\\

        2\Lambda\; e^{2\beta\phi}+2\Lambda_1\; e^{2\beta_1\phi}+2\Lambda_2\; e^{2\beta_2\phi},\;\;\;\;\;\mbox{for}\;\;\;\beta\neq1.
    \end{array} \right.\e
    where \b
    \beta_0=p(2B_1-\alpha),\;\;\;\lambda_1=\frac{p(\alpha-2)+1}{B_1\;b^{^{2pB_1}}}q^{2p}2^{p-1},\;\;\;\lambda_2=-\frac{2}{b^2},\;\;\;\lambda_3=-\lambda_1,\;\;\;\;B_1=\frac{1+\alpha
        p}{2p-1},\e \b
    \beta_1=\frac{1}{\beta},\;\;\;\beta_2=p\left(\frac{B}{\gamma}-\alpha\right),\;\;\;\;
    \Lambda_1=\frac{\beta^2}{b^2(\beta^2-1)},\;\;\;\;\Lambda_2=\frac{2^{p-1}\gamma[\alpha
        p-\beta(2 p-1)]q^{2p}}{b^{2pB}[\gamma(\alpha p+\beta)-p B]}.\e It is
    notable that the solution given by Eq.(II.18) is compatible with the
    solution obtained in my previous work \cite{IJMPD}. Also, it must be
    noted that in the absence of dilatonic field $\phi$, we have
    $V(\phi=0)=2\Lambda=-6\ell^{-2}$ and the action \eqref{2.1} reduces to
    that of Einstein-$\Lambda$-Maxwell theory \cite{ri, 4dn}.

    Now, making use of Eqs.(II.14), (II.15) and (II.18) the metric function
    $W(r)$ can be obtained as \b W(r)= \left\{\begin{array}{ll}

        -\frac{m}{r^{1-2\beta\gamma}}+(1+\beta^2)\left[\frac{1}{1-\beta^2}\left(\frac{r}{b}\right)^{2\beta\gamma}-\frac{\Lambda b^{2}(1+\beta^2)}{3-\beta^2}\left(\frac{r}{b}\right)^{\frac{2\gamma}{\beta}}
        +\frac{q^{2p}2^{p-1}\Upsilon(\beta)}{(B-1)b^{^{2(^pB-1)}}}\left(\frac{b}{r}\right)^{2\eta}\right],\;\;\mbox{for}\;\;\beta\neq1,\sqrt{3},\\\\

        -m\;r^{1/2}-2\left[\left(\frac{r}{b}\right)^{2/3}+2\Lambda(b^3r)^{\frac{1}{2}} \ln\left(\frac{r}{L}\right)
        -\frac{2^{^p}q^{2p}\Upsilon(\beta=\sqrt{3})}{({\cal{B}}-1) b^{^{2(^p{\cal{B}}-1)}}}\left(\frac{b}{r}\right)^{2\xi}\right],\;\;\;\;\mbox{for}\;\;\;\;\;\beta=\sqrt{3},\\\\

        -m+2\left[2-b^2(\Lambda+\lambda_1)+\ln\left(\frac{b}{r}\right)\right]\left(\frac{r}{b}\right)
        +\frac{p\;2^{^{p+1}}q^{2p}}{B_1(B_1-1)b^{^{2(^pB_1-1)}}}\left(\frac{b}{r}\right)^{B_1-1},\;\;\mbox{for}\;\;\beta=1,

    \end{array} \right.\e where, $m$ is an integration constant, $L$ is a dimensional constant and \b {\cal{B}}=\frac{1+\sqrt{3} \alpha
        p}{2(2p-1)},\;\;\; \eta=pB-p\alpha\gamma-1,\;\;\;\xi=p\left({\cal{B}}-\frac{\alpha\sqrt{3}}{4}\right)-1,\;\;\;
    \Upsilon(\beta)=2p-1-\frac{\beta[\alpha p-\beta\left(2p-1\right)]}{p(B-\alpha\gamma)(1+\beta^2)-\beta^2}.\e

    Note that in the case $p=1$ the metric function (II.21) is compatible
    with that of ref. \cite{IJMPD}. The plots of metric functions
    $W(r)$, presented in Eq.(II.21), have been shown in Figs.1-3 for
    $\beta=\alpha$ and $\beta\neq\alpha$ cases, separately. From the
    curves of Figs.1-3 it is understood that, for the suitably fixed
    parameters, the metric functions $W(r)$ can produce black holes with
    two horizons, extreme black holes and naked singularity black holes
    for all of $\beta\neq1,\;\sqrt{3}$, $\beta=\sqrt{3}$ and $\beta=1$
    cases.

    \begin{figure}[H]
    \centering
    \includegraphics[scale=0.7]{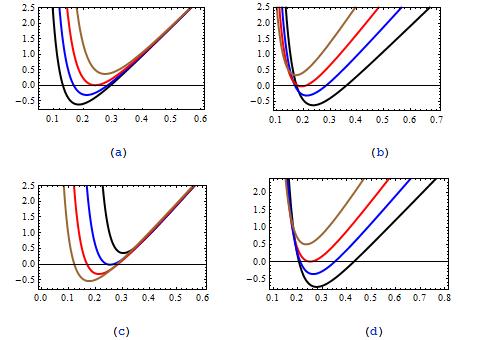}
        \caption{$W(r)$ versus $r$ for $M=1,\;Q=0.4,\;\Lambda=-3,\;b=2$ and
            $\beta\neq1,\;\sqrt{3},\;$Eq.(II.21).
            \\
            (a) $\beta=0.5,\;p=0.7$  and $\alpha=0.45,\; 0.7,\; 0.99,\; 1.3$ for
            black, blue, red and brown curves, respectively.
            \\
            (b)
            $\alpha=0.8,\;p=0.7$ and $\beta=0.48,\; 0.56,\; 0.645,\; 0.72$ for
            black, blue, red and brown curves, respectively.
            \\
            (c)
            $\alpha=0.8,\;\beta=0.6$ and $p=0.62,\; 0.666,\; 0.72,\; 0.8$ for
            black, blue, red and brown curves, respectively.
            \\
            (d) $p=0.623$ and
            $\alpha=\beta=0.4,\; 0.5,\; 0.574,\; 0.64$ for black, blue, red and
            brown curves, respectively.} \label{fig1}\end{figure}

    \begin{figure}[H]
    \centering
    \includegraphics[scale=0.7]{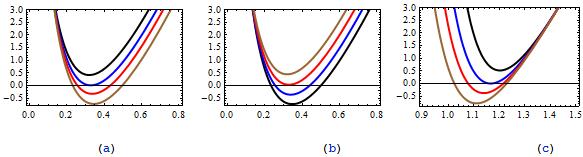}
        \caption{$W(r)$ versus $r$ for
            $M=1,\;Q=1.5,\;\Lambda=-3,\;b=1.5,\;L=1$ and
            $\beta=\sqrt{3},\;$Eq.(II.21).
            \\
            (a)
            $p=0.8$ and $\alpha=0.665,\; 0.676,\; 0.688,\; 0.7$ for black, blue,
            red and brown curves, respectively.
            \\
             (b) $\alpha=0.58$ and
            $p=0.72,\; 0.735,\; 0.748,\; 0.758$ for black, blue, red and brown
            curves, respectively.
            \\
            (c) $\alpha=\beta=\sqrt{3}$ and
            $p=0.62,\; 0.624,\; 0.628,\;0.633$ for black, blue, red and brown
            curves, respectively.} \label{fig2}\end{figure}

    \begin{figure}[H]
    \centering
    \includegraphics[scale=0.7]{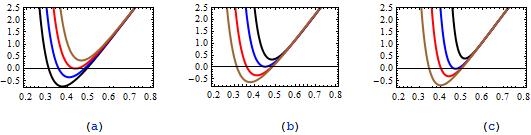}
        \caption{$W(r)$ versus $r$ for $M=2.5,\;Q=0.5,\;\Lambda=-3,\;b=1.5$
            and $\beta=1,\;$Eq.(II.21).
            \\
            (a)
            $p=0.7$ and $\alpha=0.84,\; 1.0,\; 1.18,\; 1.35$ for black, blue,
            red and brown curves, respectively.
            \\
            (b)
            $\alpha=1.3$ and $p=0.69,\; 0.718,\; 0.75,\; 0.785$ for black, blue,
            red and brown curves, respectively.
            \\
            (c) $\alpha=\beta=1$
            and $p=0.65,\; 0.675,\; 0.7,\;0.73$ for black, blue, red and brown
            curves, respectively.} \label{fig3}\end{figure}

    In order to investigate the space time singularities, one needs to
    calculate the curvature scalars. The Ricci and Kretschmann scalars,
    after some algebraic calculations, can be written in the following
    forms \b
    {\cal{R}}=ar^{2\beta\gamma-2}+b\frac{W(r)}{r^2}+c\frac{W'(r)}{r}-W''(r),\e
    \b
    {\cal{R}}^{\mu\nu\rho\lambda}{\cal{R}}_{\mu\nu\rho\lambda}=\frac{4}{r^{4-4\beta\gamma}}+a_0\frac{W(r)}{r^{4-2\beta\gamma}}+a_1\left(\frac{W(r)}{r^2}\right)^2
    +a_2\frac{W(r)W'(r)}{r^3}
    +a_3\left(\frac{W'(r)}{r}\right)^2+(W'')^2,\e where, the
    coefficients $a$, $b$, $c$ and $a_0$, $a_1$, $a_2$ and $a_3$ are
    functions of dilaton and nonlinearity parameters. Now, making use of
    the metric function (II.21) in Eqs.(II.23) and (II.24), one can show
    that the Ricci and Kretschmann scalars are finite for finite values
    of $r$, and
    \b\lim_{\;\;\;\;\;\;\;\;\;\;\;\;r\longrightarrow\infty}{\cal{R}}=0,\;\;\;\;\;\;\;\;\;\;\mbox{and}\;\;\;\;\;\;\;\;\;\;\lim_{\;\;\;\;\;\;\;\;\;\;\;\;r\longrightarrow0}{\cal{R}}=\infty,\e
    \b\lim_{\;\;\;\;\;\;\;\;\;\;\;\;r\longrightarrow\infty}{\cal{R}}^{\mu\nu\rho\lambda}{\cal{R}}_{\mu\nu\rho\lambda}=0,\;\;\;\;\;\;\;\;\;\;\mbox{and}\;\;\;\;\;\;\;\;\;\;\lim_{\;\;\;\;\;\;\;\;\;\;\;\;r\longrightarrow0}{\cal{R}}^{\mu\nu\rho\lambda}{\cal{R}}_{\mu\nu\rho\lambda}=\infty\e
    Equations (II.25) and (II.26) show that there is an essential
    singularity located at $r=0$ and the asymptotic behavior of the
    solutions is neither flat nor AdS. It means that inclusion of the
    scalar field modifies the asymptotic behavior of the solutions.

    It is well-known that existence of at least one event horizon and
    appearance of the curvature singularities are two necessary
    conditions to be satisfied simultaneously, in order to the solutions
    be interpreted as the black holes \cite{4dRG, bh, es}. The plots of
    Figs.1-3 and Eqs.(II.25) and (II.26) show that both of these
    requirements are fulfilled by the solutions obtained here. As the
    result, our new solutions are really black holes.

    \setcounter{equation}{0}
    \section{Black hole thermodynamics}

    The aim of this section is to check the validity of the first law of
    black hole thermodynamics for all of the new charged dilatonic black
    holes obtained in the previous section. For this purpose, we
    calculate the conserved and thermodynamic quantities related to
    either of the black hole solutions. The black hole entropy as a pure
    geometrical quantity can be obtained from the well-known
    entropy-area law. According to this nearly universal law, which is
    valid in the Einstein-dilaton gravity theory, the black hole entropy
    is equal to one quarter of the black hole surface area and for our
    new black hole solutions can be written in the following form
    \cite{kich, myu} \b S=\frac{A}{4}=\pi
    b^2\left(\frac{b}{r_+}\right)^{2\beta\gamma-2},\e which reduces to
    the entropy relation of the Einstein black holes in the absence of
    dilatonic parameters ($\beta=0=\gamma$).

    The other thermodynamical quantity which can be calculated
    geometrically is the Hawking temperature associated with the black
    hole horizon $r=r_+$. In terms of the surface gravity $\kappa$ it
    can be written as $T=\frac{\kappa}{2\pi}$, with
    $\kappa=\sqrt{-\frac{1}{2}(\nabla_\mu\chi_\nu)(\nabla^\mu\chi^\nu)}=\frac{1}{2}W'(r_+)$,
    and $\chi^\mu$ is null killing vector of the horizon. After some algebraic calculations we arrived at \cite{log, myu}
    \b T=\frac{1}{4\pi r_+}\left\{\begin{array}{ll}
        (1+\beta^2)\left[\frac{1}{1-\beta^2}\left(\frac{r_{+}}{b}\right)^{2\beta\gamma}
        -\Lambda b^2\left(\frac{b}{r_{+}}\right)^{2(\beta\gamma-1)}
        -\frac{2^{^{p-1}}q^{2p}\Upsilon(\beta)}{b^{^{2(pB-1)}}}\left(\frac{b}{r_{+}}\right)^{2\eta}\right],\beta\neq\sqrt{3},1,\\\\

        -2\left(\frac{b}{r_{+}}\right)^{-\frac{3}{2}}-4\Lambda b^2\left(\frac{b}{r_{+}}\right)^{-\frac{1}{2}}
        -\frac{2^{^{p+1}}q^{2p}\Upsilon(\beta=\sqrt{3})}{b^{^{2(^p{\cal{B}}-1)}}}\left(\frac{b}{r_+}\right)^{2\xi},\;\;\;\;\;\;\;\;\;\beta=\sqrt{3},\\\\

        \frac{2r_{+}}{b}\left[1-b^2(\Lambda+\lambda_1)+\ln\left(\frac{b}{r_+}\right)-\frac{p\;2^{^{p+2}}q^{2p}}{B_1b^{^{2(^pB_1-1)}}}\left(\frac{b}{r_{+}}\right)^{B_1}\right],
        \;\;\;\;\;\;\;\;\;\beta=1. \end{array} \right.\e

    The black hole temperature (III.2) reduces to that of ref. \cite{IJMPD}, if one let $p=1$. Note
    that we have used the relation $W(r_+)=0$ for eliminating the mass
    parameter $m$ from the obtained equations. Also, it must be noted
    that extreme black holes occur if $q=q_{ext}$ and $r_+=r_{ext}$ be
    chosen such that $T=0$. With this issue in mind and making use of
    Eq.(III.2) one can show that the extreme black holes exist if the
    following equations are satisfied \b \frac{1}{1-\beta^2}
    -\Lambda r_{ext}^2\left(\frac{b}{r_{ext}}\right)^{4\beta\gamma}
    -\frac{2^{^{p-1}}q_{ext}^{2p}\Upsilon(\beta)}{b^{^{2(pB-1)}}}\left(\frac{b}{r_{ext}}\right)^{2(\eta+1)}=0,\;\; \mbox{for}\;\;\beta\neq\sqrt{3},1,\e

    \b 1+2\Lambda b r_{ext}
    +\frac{2^{^{p}}q_{ext}^{2p}\Upsilon(\beta=\sqrt{3})}{b^{^{2(^p{\cal{B}}-1)}}}\left(\frac{b}{r_{ext}}\right)^{{\cal{B}}-2}=0,\;\;\;\;\;\; \mbox{for}\;\;\;\;\;\;\beta=\sqrt{3},\e

    \b
    1-b^2(\Lambda+\lambda_1)+\ln\left(\frac{b}{r_{ext}}\right)-\frac{p\;2^{^{p+2}}q_{ext}^{2p}}{B_1b^{^{2(^pB_1-1)}}}\left(\frac{b}{r_{ext}}\right)^{B_1}=0,\;\;
    \mbox{for}\;\;\beta=1.\e

    In order to investigate the effects of dilatonic and nonlinearity
    parameters (i.e. $\alpha,\;\beta\;$ and $p$) on the horizon
    temperature of the black holes, the plots of black hole temperature
    versus horizon radius have been shown in Figs. 4-6 by considering
    the $\alpha=\beta$ and $\alpha\neq\beta$ cases, separately. The
    plots of Figs. 4 and 6 show that, in the cases $\beta\neq\sqrt{3},1$
    and $\beta=1$, the extreme black holes can occur for $r_+=r_{ext}$
    only. Also, the physical black holes with positive temperature are
    those for which $r_+>r_{ext}$ and un-physical black holes, having
    negative temperature, occur if $r_+<r_{ext}$. Regarding the plots of
    Fig. 5 (for black holes with $\beta=\sqrt{3}$) one can argue that
    the equation $T=0$ have two real roots located at $r_+=r_{1ext}$ and
    $r_+=r_{1ext}$, where the extreme black holes can occur. The black
    holes are physically acceptable if their horizon radii be in the
    range $r_{1ext}<r_+<r_{2ext}$.

    The electric potential $\Phi$ of black holes, measured by an
    observer located at infinity with respect to the horizon, can be
    calculated making use of the following standard relation
    \cite{stete, bi3, 92, 93} \b
    \Phi=A_\mu\chi^\mu|_{\mbox{reference}}-A_\mu\chi^\mu|_{r=r_+},\e
    here, $\chi=C\partial_t$ is the null generator of the horizon and
    $C$ is an arbitrary constant to be determined \cite{he, kz1}. Noting
    Eqs.(II.14) and (III.6) we obtained the black hole's electric potential
    on the horizon as \b \Phi=\frac{C q}{B-1}r_+^{1-B}.\e The conserved
    electric charge of the black holes can be obtained by calculating
    the total electric flux measured by an observer located at infinity
    with respect to the horizon (i.e. $r\rightarrow\infty$) \cite{es,
        pa2}. It can be obtained with the help of Gauss's electric law which
    can be written as \cite{he, kord} \b
    Q=\frac{1}{4\pi}\int_{r\rightarrow\infty}r^2e^{2\beta\phi(r)}(-{\cal{F}})^{p-1}e^{-2p\alpha\phi(r)}F_{\mu\nu}u^\mu
    u^\nu d\Omega,\e where $u_\mu$ and $u_\nu$ are timelike and
    spacelike unit vectors normal to the hypersurface of radius $r$,
    respectively. Making use of Eqs.(II.2), (II.13) and (II.14), after some
    simple calculations, we arrived at \b
    Q=2^{^{p-1}}b^2\left(\frac{q}{b^B}\right)^{2p-1},\e  which is
    compatible with the results of our previous work in the case $p=1$
    \cite{IJMPD}. It reduces to the charge of
    Reissner-Nordstr$\ddot{\mbox{o}}$m-A(dS) black holes in the absence
    of dilatonic field. Also, a redefinition of the integration constant
    $q$ makes this relation consistent with the result of
    refs.\cite{stete, kord}.

    The other conserved quantity to be calculated is the black hole
    mass. As mentioned before, it can be obtained in terms of the mass
    parameter $m$. Since the asymptotic behavior of the metric functions
    given by Eq.(II.21) is unusual, the Brown and York quasilocal
    formalism must be used for obtaining the quasilocal mass \cite{man1,
        man11}. By considering a metric in the following form (Eq.(II.7) in
    ref.\cite{man2}) \b
    ds^2=-X^2(\rho)dt^2+\frac{d\rho^2}{Y^2(\rho)}+\rho^2\left(d\theta^2+\sin^2\theta
    d\varphi^2\right),\e provided that the matter field does not contain
    derivatives of the metric, the quasilocal black hole mass can be
    obtained through the following relation (Eq.(II.8) in
    ref.\cite{man2}) \b{\cal{M}}=\rho X(\rho)[Y_0(\rho)-Y(\rho)],\e in
    which $Y_0(\rho)$ is a background metric function which determines
    the zero of the mass.

    In order to obtain the analogous Arnowitt-Deser-Misner (ADM) mass
    $M$, the limit $\rho\rightarrow\infty$ must be taken \cite{man2}.
    Now, we must to write the metric (II.6) in the form of Eq.(III.10).
    This can be done by considering the transformation $\rho=rR(r)$,
    from which on can show that
    $$ dr^2=\frac{d\rho^2}{(1-\beta\gamma)^2R^2(r)}.$$ Therefore, in
    our case, we have \b X^2(\rho)=W\left(r\right),\;\;\;\;\;\;\;
    \mbox{and}\;\;\;\;\;\;\;\;\;\;\;\;\;\;\;\;\;
    Y^2(\rho)=\frac{R^2(r)}{(1+\beta^2)^2}W\left(r\right),\;\;\;\;\;\;\;\;
    \mbox{with}\;\;\;\;\;\;\;\; r=r(\rho).\e By substituting these
    quantities into Eq.(III.11) and taking the limit
    $\rho\rightarrow\infty$ or equivalently $r\rightarrow\infty$ the
    total mass of the charged dilatonic black holes, identified here, is
    obtained as (see appendix-A) \b
    M=\frac{m\;b^{2\beta\gamma}}{2(1+\beta^2)},\e which is compatible
    with the result of refs.\cite{ko3, IJMPD, stete}. Also, it recovers
    the mass of Reissner-Nordstr$\ddot{\mbox{o}}$m-A(dS) black holes
    when the dilatonic potential disappears.

    Now, we are in the position to investigate the consistency of these
    quantities with the thermodynamical first law. From Eqs. (II.21),
    (III.1), (III.9) and (III.13), we can obtain the black hole mass as the
    function of extensive parameters $S$ and $Q$. To do so, we use the
    relation $W(r_+)=0.$ The Smarr-type mass formula for the new black
    holes can be obtained as \b M(r_+,q)=\left\{\begin{array}{ll}
        \frac{b}{2} \left[\frac{1}{1-\beta^2}\frac{r_+}{b}-\frac{\Lambda
            b^2(1+\beta^2)}{3-\beta^2}\left(\frac{b}{r_+}\right)^{4\beta\gamma-3}
        +\frac{\;2^{^{p-1}}q^{2p}\Upsilon(\beta)}{(B-1)b^{^{2(pB-1)}}}\left(\frac{b}{r_+}\right)^{2\eta+2\beta\gamma-1}\right],\;\;\beta\neq1,\;\sqrt{3},\\\\

        -\frac{b}{4}\left[\frac{r_+}{b}+2\Lambda b^2\ln\left(\frac{r_+}{L}\right)
        -\frac{2^{^{p}}q^{2p}\Upsilon(\beta=\sqrt{3})}{({\cal{B}}-1)b^{^{2(p{\cal{B}}-1)}}}
        \left(\frac{b}{r_+}\right)^{\frac{1}{2}(\sqrt{3}\alpha-1)}\right],\;\;\;\;\;\;
        \beta=\sqrt{3},\\\\

        \frac{b}{2}\left\{\left[2-b^2(\Lambda+\lambda_1)+\ln\left(\frac{b}{r_{+}}\right)\right]\left(\frac{r_+}{b}\right)
        +\frac{p\;2^p\;q^{2p}}{B_1(B_1-1)b^{^{2(pB_1-1)}}}\left(\frac{b}{r_+}\right)^{B_1-1}\right\},
        \;\;\beta=1.\end{array} \right.\e

    It is a matter of calculation to show that the intensive parameters
    $T$ and $\Phi$, conjugate to the black hole entropy and charge,
    satisfy the following relations \b \left(\frac{\partial M}{\partial
        S}\right)_Q=T, \;\;\;\;\; \mbox{and}\;\;\;\;\; \left(\frac{\partial
        M}{\partial Q}\right)_S=\Phi,\e provided that $C$ be chosen as
    \cite{kz1}

    \b C=\left\{\begin{array}{ll} \frac{p(2-\Upsilon_1)}{1+\alpha p}\;\;\;\mbox{for}\;\;\;\beta=1,\\\\

        \frac{p\Upsilon(\beta)}{2p-1}\;\;\;\;\;\;\mbox{for}\;\;\;\;\;\;\beta\neq1.
    \end{array} \right.\e
    Note that $\Upsilon_1=(\alpha p-2p+1)(\alpha p-2p+2)(2p-1)^{-1}$ and
    the condition $r_+=b$ has been used for the case $\beta=1$. Also,
    Eq.(III.16) reduces to its corresponding value in the
    Einstein-Maxwell-dilaton gravity theory \cite{IJMPD}. Therefore, we
    proved that the first law of black hole thermodynamics is valid, for
    all of the new nonlinearly charged dilatonic black holes, in the
    following form \b d M(S,Q)= T dS+\Phi dQ.\e Here, $S$ and $Q$ are
    known as the thermodynamical extensive parameters and $T$ and $\Phi$
    are intensive parameters conjugate to $S$ and $Q$, respectively.
    From Eq.(III.17) one can argue that even if the conserved and
    thermodynamic quantities are affected by dilaton and nonlinearity
    parameters, the first law of black hole thermodynamics remains
    valid.

    \setcounter{equation}{0}
    \section{Black hole local stability}

    In this section, we investigate the thermal stability or phase
    transition of our new the black hole solutions, making use of the
    canonical ensemble method. To do so, we need to calculate the black
    hole heat capacity with the black hole charge as a constant. It is
    defined in the following form
    \b C_Q =T\left(\frac{\partial
        S}{\partial T}\right)_Q= \frac{T}{M_{SS}}.\e The last step in
    Eq.(IV.1) comes from the fact that $T=\left(\partial M/\partial
    S\right)_Q$ and we have used the definition $M_{SS}=
    \left(\partial^2M/\partial S^2\right)_Q$.

    It is well-known that, the positivity of the black hole heat
    capacity $C_Q$ or equivalently the positivity of $(\partial
    S/\partial T)_Q$ or $M_{SS}$ is sufficient to ensure the local
    stability of the physical black holes. The unstable black holes
    undergo phase transitions to be stabilized. The sign of the
    black hole heat capacity changes, from negative to positive, at its vanishing points. Thus,
    they signal the existence of a kind of phase transition. In
    addition, an unstable black hole undergoes phase transition at the divergent points of the black hole heat
    capacity where the denominator of the heat capacity vanishes.
    These two kinds of thermodynamic phase transitions are called as type-1 and type-2
     phase transitions, respectively \cite{mh1, mh2, mh3} (see also \cite{4dRG, myu, 3dst}). Considering the above mentioned
    points, we proceed to perform a thermal stability or phase
    transition analysis for all of the new black hole solutions we just
    obtained.

    \subsection{Black holes with $\beta\neq1,\;\sqrt{3}$}

    Making use of Eq.(III.14) and noting Eq.(III.1), the denominator of the
    black hole heat capacity can be calculated as \b
    M_{SS}=\frac{-(1+\beta^2)}{8\pi^2
        b^3}\left(\frac{b}{r_+}\right)^{1-2\beta\gamma}\left[\left(\frac{b}{r_+}\right)^{2-2\beta\gamma}
    +\Lambda b^{2}(1-\beta^2)\left(\frac{b}{r_+}\right)^{2\beta\gamma}
    -\frac{2^{^{p-1}}q^{2p}\Upsilon(\beta)}{b^{^{2(pB-1)}}}\left(\frac{b}{r_+}\right)^{2\eta+2}\right].\e
    The real roots of equation $M_{SS}=0$ indicate the points of type-2
    phase transition. As it is too difficult to obtain the real roots of
    this equation analytically, we have plotted $M_{SS}$ versus $r_+$ in
    Fig. 4 for different values of dilaton and nonlinearity parameters.
    The plots show that, for the properly fixed parameters, $M_{SS}$ is
    positive valued every where and no type-2 phase transition can take
    place. This kind of black holes undergo type-1 phase transition at
    the point $r_+=r_{ext}$ where the black hole heat capacity vanishes.
    The black holes  with the horizon radii in the range $r_+>r_{ext}$
    are locally stable.
    \begin{figure}[H]
    \centering
    \includegraphics[scale=0.7]{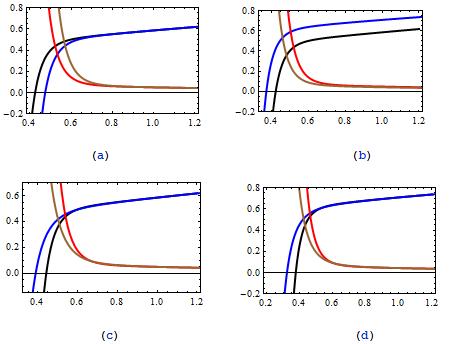}
        \caption{$T$ and $M_{SS}$ versus $r_+$ for
            $Q=0.4,\;\Lambda=-3,\;b=2$ and $\beta\neq1,\;\sqrt{3},\;$Eqs.(III.2)
            and (IV.2).
            \\
            (a) $\beta=0.6,\;p=0.7$ and
            $[T:\alpha=0.7(\mbox{black}),\; 1.0(\mbox{blue})]$ and
            $[2M_{SS}:\alpha=0.7(\mbox{red}),\; 1.0(\mbox{brown})].$
            \\
            (b) $\alpha=0.7,\;p=0.6$ and $[T:\beta=0.6(\mbox{black}),\;
            0.7(\mbox{blue})]$ and $[2M_{SS}:\alpha=0.6(\mbox{red}),\;
            0.7(\mbox{brown})].\;\;\;\;\;$
            \\(c) $\alpha=0.8,\;\beta=0.6$ and
            $[T:p=0.6(\mbox{black}),\; 0.65(\mbox{blue})]$ and
            $[2M_{SS}:p=0.6(\mbox{red}),\; 0.65(\mbox{brown})].\;\;\;\;\;\;\;$
            \\
            (d) $\alpha=\beta=0.7$ and $[T:p=0.6(\mbox{black}),\;
            0.65(\mbox{blue})]$ and $[2M_{SS}:p=0.6(\mbox{red}),\;
            0.65(\mbox{brown})]$.} \label{fig4}\end{figure}

    \subsection{Black holes with $\beta=\sqrt{3}$}
    The numerator of these the black holes is given by Eq.(III.2). Also,
    it is a matter of calculation to show that its denominator is given
    by the following equation \b M_{SS}=\frac{-1}{2
        \pi^2b^3}\left[1-2\Lambda b^{2}\left(\frac{b}{r_+}\right)
    -\frac{2^{^{p}}q^{2p}(2{\cal{B}}-1)\Upsilon(\beta=\sqrt{3})}{b^{^{2(p{\cal{B}}-1)}}}\left(\frac{b}{r_+}\right)^{{\cal{B}}}\right].\e
    The plots of denominator together with the numerator of the black
    hole heat capacity are shown in Fig. 5 for different values of
    nonlinearity and dilaton parameters. Regarding the plots of Fig.5,
    it is easily understood that there are two points of type-1 phase
    transition located at the points $r_+=r_{1ext}$ and $r_+=r_{2ext}$
    with $r_{1ext}<r_{2ext}$. The black hole heat capacity diverges at
    the real root of $M_{SS}=0$ which appears at the point $r_+=r_{0}$
    with $r_{0}<r_{1ext}$. Also, the physical black holes, those having
    positive temperature, are unstable every where. Because they occur
    in the range $r_{1ext}<r_+<r_{2ext}$ and their heat capacity is
    negative in this range.

    \begin{figure}[H]
    \centering
    \includegraphics[scale=0.7]{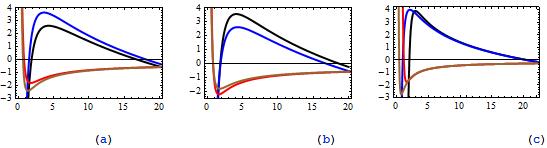}
    \caption{$T$
            and $M_{SS}$ versus $r_+$ for $Q=2.5,\;\Lambda=-3,\;b=1.5$ and
            $\beta=\sqrt{3},\;$Eqs.(III.2) and (IV.3).$\;\;\;\;\;\;\;\;\;$
            \\
            (a)
            $p=0.8$ and $[10T:\alpha=0.8(\mbox{black}),\; 0.9(\mbox{blue})]$ and
            $[20M_{SS}:\alpha=0.8(\mbox{red}),\;
            0.9(\mbox{brown})].\;\;\;\;\;\;\;\;\;\;\;\;\;\;\;\;\;\;\;\;$
            \\
            (b)
            $\alpha=0.8$ and $[10T:p=0.75(\mbox{black}),\; 0.8(\mbox{blue})]$
            and $[20M_{SS}:p=0.75(\mbox{red}),\;
            0.8(\mbox{brown})].\;\;\;\;\;\;\;\;\;\;\;\;\;\;\;$
            \\
            (c)
            $\alpha=\beta=\sqrt{3}$ and $[5T:p=0.6(\mbox{black}),\;
            0.8(\mbox{blue})]$ and $[10M_{SS}:p=0.6(\mbox{red}),\;
            0.8(\mbox{brown})].$} \label{fig5}\end{figure}

    \subsection{Black holes with $\beta=1$}

    Starting from Eq.(III.14) and using Eq.(III.1), one can calculate the
    denominator of the black hole heat capacity. It can be written in
    the following form \b M_{SS}=\frac{-1}{2\pi^2
        b^2r_+}\left[1-\frac{p\;2^{^{p}}q^{2p}}{b^{^{2(pB_1-1)}}}\left(\frac{b}{r_+}\right)^{B_1}\right].\e
    In order to investigate the points of type-1 and type-2 phase
    transitions and to determine the ranges at which the black hole heat
    capacity is positive valued, we have plotted the numerator and
    denominator of the black hole heat capacity in Fig. 6. The plots
    show that there is a point of type-1 phase transition located at
    $r_+=r_{ext}$, where the black hole heat capacity vanishes. The
    black hole heat capacity diverges at $r_+=r_1$ and it is point of
    type-2 phase transition. This kind of black holes are locally stable
    provided that the condition $r_{ext}<r_+<r_{1}$ is fulfilled.

    \begin{figure}[H]
    \centering
    \includegraphics[scale=0.7]{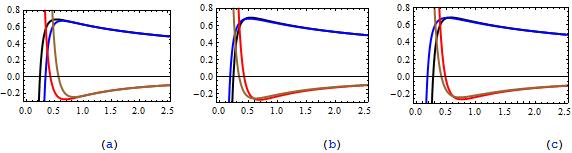}
        \caption{$T$ and $M_{SS}$ versus $r_+$ for
            $Q=0.4,\;\Lambda=-3,\;b=2$ and $\beta=1,\;$Eqs.(III.2) and
            (IV.4).$\;\;\;\;\;\;\;\;\;\;\;\;\;\;\;$
            \\
            (a) $p=0.7$ and
            $[T:\alpha=0.7(\mbox{black}),\; 1.5(\mbox{blue})]$ and
            $[10M_{SS}:\alpha=0.7(\mbox{red}),\;
            1.5(\mbox{brown})].\;\;\;\;\;\;\;\;\;\;\;\;\;\;\;\;\;\;\;\;$
            \\
            (b)
            $\alpha=0.7$ and $[T:p=0.7(\mbox{black}),\; 0.8(\mbox{blue})]$ and
            $[10M_{SS}:p=0.7(\mbox{red}),\;
            0.8(\mbox{brown})].\;\;\;\;\;\;\;\;\;\;\;\;\;\;\;\;\;\;\;\;$
            \\
            (c)
            $\alpha=\beta=1$ and $[T:p=0.7(\mbox{black}),\; 0.8(\mbox{blue})]$
            and $[10M_{SS}:p=0.7(\mbox{red}),\; 0.8(\mbox{brown})].$}
        \label{fig6}\end{figure}


    \setcounter{equation}{0}
    \section{Black hole global stability}

    Study of the black hole global stability was initially proposed by
    Hawking and Page, as the pioneers of this idea \cite{hp}. Based on
    this proposal, one can investigate the global stability of black
    holes by studying the corresponding Gibbs free energy. The Gibbs
    free energy of the charged black holes in the grand canonical
    ensemble is given by \cite{grg} \b G=M-TS-Q\Phi.\e The Gibbs free
    energy is required to be positive to ensure global stability of the
    black holes with positive temperature. The Hawking-Page phase
    transition can occur at the points where the Gibbs free energy
    vanishes. The black hole temperature at which the Hawking-Page phase
    transition takes place is dubbed as the critical temperature
    ($T_H$). At this temperature, Hawking-Page phase transition between
    black hole and thermal state (radiation) occurs\cite{hp, grg, plb}.
    In the following subsections we calculate the Gibbs free energy of
    our new black holes and, regarding the above mentioned points,
    analyze their global stability.

    \subsection{Black holes with $\beta\neq1,\;\sqrt{3}$}

    Noting Eqs.(III.1), (III.2), (III.7), (III.9), (III.14) and (V.1), after some
    algebraic calculations we arrive the Gibbs free energy as follows
    $$ G=\frac{r_+}{4}+\frac{\Lambda
        b^3(1-\beta^4)}{4(3-\beta^2)}\left(\frac{b}{r_+}\right)^{4\beta\gamma-3}
    -\frac{\;2^{^{p-1}}q^{2p}\Upsilon(\beta)r_+}{(B-1)b^{^{2(pB-1)}}}\left[\frac{p}{2p-1}\left(\frac{b}{r_+}\right)^{B}\right.$$
    \b \left.-\frac{1}{4}\left\{1-\beta^2+B(1+\beta^2)\right\}\left(\frac{b}{r_+}\right)^{2\eta+2\beta\gamma}\right].\e
    We need the real roots of equation $G(r_+)=0$, but this is a
    difficult task analytically. So we plot the curves of $G$ in terms
    of $r_+$. They are shown in Fig. 7. It is seen from Fig. 7 that at
    the vanishing point of Gibbs free energy labeled by $r_+=r_c$ the
    Hawking-Page phase transition occurs.  For $r_{ext}<r_+<r_c$, where
    both the Gibbs free energy and temperature are positive the black
    hole is globally stable. In $r_+=r_{ext}$ the temperature is zero
    (extremal black hole) but in $r_+=r_c$ temperature is equal to $T_H$
    at which the Gibbs free energy vanishes. We should mention that in
    the range $r_{ext}<r_+<R_c$ the Gibbs free energy is a decreasing
    function of $r_+$. So the larger black holes are more stable ones.

    \begin{figure}[H]
    \centering
    \includegraphics[scale=0.7]{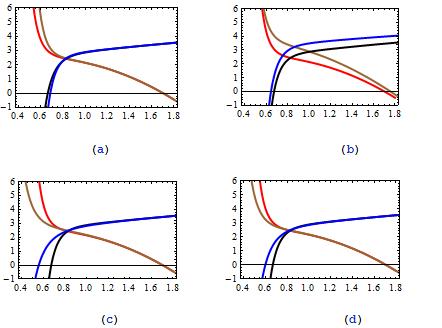}
        \caption{$T$ and $G$ versus $r_+$ for $q=1,\;\Lambda=-3,\;b=2$ and
            $\beta\neq1,\;\sqrt{3},\;$Eqs.(III.2) and
            (V.2).$\;\;\;\;\;\;\;\;\;\;\;\;$
            \\
            (a) $\beta=0.6,\;p=0.6$ and
            $[5T:\alpha=0.4(\mbox{black}),\; 1.2(\mbox{blue})]$ and
            $[G:\alpha=0.4(\mbox{red}),\;
            1.2(\mbox{brown})].\;\;\;\;\;\;\;\;\;\;$
            \\
            (b) $\alpha=0.8,\;p=0.6$
            and $[5T:\beta=0.6(\mbox{black}),\; 0.8(\mbox{blue})]$ and
            $[G:\beta=0.6(\mbox{red}),\;
            0.8(\mbox{brown})].\;\;\;\;\;\;\;\;\;\;$
            \\
            (c)
            $\alpha=0.8,\;\beta=0.6$ and $[5T:p=0.6(\mbox{black}),\;
            0.7(\mbox{blue})]$ and $[G:p=0.6(\mbox{red}),\;
            0.7(\mbox{brown})].\;\;\;\;\;\;\;\;\;\;$
            \\
            (d) $\alpha=\beta=0.6$ and
            $[5T:p=0.6(\mbox{black}),\; 0.65(\mbox{blue})]$ and
            $[G:p=0.6(\mbox{red}),\; 0.65(\mbox{brown})]$.}
        \label{fig7}\end{figure}

    \subsection{Black holes with $\beta=\sqrt{3}$}

    By use of the expressions presented in Eqs.(III.1), (III.2), (III.7),
    (III.9) and (III.14) into Eq.(V.1) we obtain the Gibbs free energy of
    the black holes correspond to the case $\beta=\sqrt{3}$. That is
    $$ G=\frac{r_+}{4}+\Lambda
    b^3\left\{1-\frac{1}{2}\ln\left(\frac{r_+}{L}\right)\right\}
    +\frac{2^{^{p-2}}q^{2p}\Upsilon(\beta=\sqrt{3})}{({\cal{B}}-1)b^{^{2(p{\cal{B}}-1)}}}r_+
    \left[\left(\frac{b}{r_+}\right)^{\frac{1}{2}(\sqrt{3}\alpha+1)}\right.$$
    \b \left. +2({\cal{B}}-1)\left(\frac{b}{r_+}\right)^{2\xi+\frac{3}{2}}-\frac{2p}{2p-1}\left(\frac{b}{r_+}\right)^{{\cal{B}}}\right].\e
    As it is difficult to solve the equation $G(r_+)=0$ and obtain its
    real roots analytically, we have shown the plots of $G$ and $T$
    versus $r_+$ in Fig.8.

    \begin{figure}[H]
    \centering
    \includegraphics[scale=0.7]{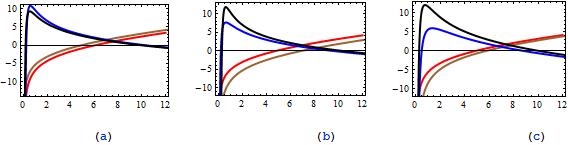} \caption{$T$ and
            $G$ versus $r_+$ for $q=1.2,\;\Lambda=-3,\;b=1.2$ and
            $\beta=\sqrt{3},\;$Eqs.(III.2) and (V.3).
            \\
             (a)
            $p=0.75$ and $[10T:\alpha=0.6(\mbox{black}),\; 1.0(\mbox{blue})]$
            and $[G:\alpha=0.6(\mbox{red}),\;
            1.0(\mbox{brown})].$
            \\
             (b)
            $\alpha=0.7$ and $[10T:p=0.65(\mbox{black}),\; 0.8(\mbox{blue})]$
            and $[G:p=0.65(\mbox{red}),\;
            0.8(\mbox{brown})].$
            \\
            (c)
            $\alpha=\beta=\sqrt{3}$ and $[10T:p=1.0(\mbox{black}),\;
            1.2(\mbox{blue})]$ and $[G:p=1(\mbox{red}),\; 1.2(\mbox{brown})].$}
        \label{fig8}\end{figure} The plots of Fig. 8 show that the Gibbs
    free energy of this class of black holes vanishes at $r_+=r_{c1}$
    and black holes with horizon radius equal to $r_{c1}$ experience
    Hawking-Page phase transition. The black holes with horizon radius
    in the range $r_{c1}<r_+<r_{2ext}$ are globally stable. Since the
    Gibbs free energy is an increasing function of $r_+$, the smaller
    black holes with the horizon radius in this range are more stable.
    For the black holes with the horizon radius smaller than $r_{c1}$
    the thermal/ or radiation state is preferred.

    \subsection{Black holes with $\beta=1$}

    Through combination of Eqs.(III.1), (III.2), (III.7), (III.9), (III.14) and
    (V.1) one is able to show that the Gibbs free energy of this kind of
    black holes, as the function of black hole horizon radius, can be
    written in the following form \b G=\frac{r_+}{2}+
    +\frac{p\;2^{p-1}\;q^{2p}r_+}{B_1(B_1-1)b^{^{2(pB_1-1)}}}\left\{4B_1-3-\frac{(2-\Upsilon_1)B_1}{(1+\alpha p)} \right\}\left(\frac{b}{r_+}\right)^{B_1}.\e
    For the purpose of global stability analysis of the black holes we
    have plotted $G$ and $T$ versus $r_+$ in Fig.9.
    \begin{figure}[H]
        \centering
        \includegraphics[scale=0.7]{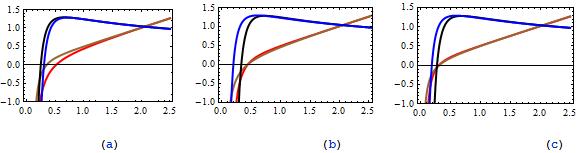}
    \caption{$T$
    and $G$ versus $r_+$ for $q=0.4,\;\Lambda=-3,\;b=1$ and
    $\beta=1,\;$Eqs.(III.2) and (V.4).
    \\
    (a) $p=0.75$ and $[2T:\alpha=0.5(\mbox{black}),\; 0.9(\mbox{blue})]$
    and $[G:\alpha=0.5(\mbox{red}),\;
    0.9(\mbox{brown})].$
    \\
    (b)
    $\alpha=0.7$ and $[2T:p=0.7(\mbox{black}),\; 0.85(\mbox{blue})]$ and
    $[G:p=0.7(\mbox{red}),\;
    0.85(\mbox{brown})].$
    \\
    (c) $\alpha=\beta=1$ and $[2T:p=0.8(\mbox{black}),\;
    0.95(\mbox{blue})]$ and $[G:p=0.8(\mbox{red}),\;
    0.95(\mbox{brown})].$}
\label{fig9}
    \end{figure}

    The plots of Fig.9 show that there is a critical radius at which
    Gibbs free energy vanishes which we label by $r_{c2}$. The black
    holes with horizon radius equal to $r_{c2}$ undergo Hawking-Page
    phase transition. This kind of black holes with the horizon radius
    greater than $r_{c2}$ are globally stable. The Gibbs free energy is
    an increasing function of $r_+$, therefore the black holes with
    smaller horizon radius are more stable. In addition the black holes
    with the horizon radius in the range $r_{ext}<r_+<r_{c2}$ prefer the
    thermal state.

    \setcounter{equation}{0}
    \section{Conclusion}

    In this work, we have studied thermodynamic properties of the new
    nonlinearly charged dilatonic four-dimensional black holes, as the
    exact solutions to the to the field equations of the
    Einstein-power-Maxwell-dilaton gravity theory. The explicit form of
    the coupled scalar, electromagnetic and gravitational field
    equations have been obtained by varying the action of
    Einstein-dilaton gravity coupled to power-Maxwell invariant as the
    matter field. By introducing a static and spherically symmetric
    geometry, we found that the solution of the scalar field equation
    can be written in the form of a generalized Liouville dilatonic
    potential. Also, three new classes of charged dilaton black hole
    solutions have been obtained in the presence of the power law
    nonlinear electrodynamics. Regarding the Ricci and Kretschmann
    scalars, we found that there is a point of essential singularity
    located at the origin. Also, the asymptotic behavior of the
    solutions are neither flat nor AdS. The existence of the real roots
    of the metric functions together with the singular Ricci and
    Kretschmann scalars are sufficient to interpret the solutions as
    black hole. Plots of Figs. 1-4, show that the new black hole
    solutions can provide two horizon, extreme and naked singularity
    black holes for the suitably fixed parameters.

    Next, we studied the thermodynamics of the new black hole solutions.
    We have obtained the conserved charge and mass of the black holes.
    Also, by using the geometrical methods, we have calculated the
    temperature, entropy and electric potential for all of the new black
    hole solutions. We showed that, for the black hole solutions with
    $\beta\neq1,\; \sqrt{3}$ and $\beta=1$ the extreme, physical and
    un-physical black holes can occur if $r_+=r_{ext}$, $r_+>r_{ext}$
    and $r_+<r_{ext}$, respectively while, for those with
    $\beta=\sqrt{3}$ the extreme black holes occur at the points
    $r_+=r_{1ext}$ and $r_+=r_{2ext}$. The radii of physical black holes
    are in the range $r_{1ext}<r_+<r_{2ext}$ and un-physical black holes
    are in the ranges $r_+<r_{1ext}$ and $r_+>r_{2ext}$. Through a
    Smarr-type mass formula, we have obtained the black hole mass as the
    function of the thermodynamical extensive parameters $S$ and $Q$,
    from which we have obtained the intensive parameters $T$ and $\Phi$.
    Compatibility of the results obtained from thermodynamical and
    geometrical approaches proves the validity of the thermodynamical
    first law for all of the new black hole solutions.

    Then, from the canonical ensemble point of view, we have analyzed
    the thermal stability or phase transition of the new black hole
    solutions. Regarding the signature of the black hole heat capacity
    with the black hole charge as a constant, we found that the
    following possibilities are considerable. (\emph{I}) For the heat
    capacity of the black holes with $\beta\neq1,\;\sqrt{3}$ there is no
    divergent point and no type-2 phase transition occur. Type-1 phase
    transition takes place at the point $r_+=r_{ext}$ where the black
    hole heat capacity vanishes. This class of black holes remain stable
    for the horizon radii in the range $r_+>r_{ext}$ (Fig. 4).
    (\emph{II})The black holes corresponding to $\beta=\sqrt{3}$  have
    two points of type-1 phase transition labeled by $r_+=r_{1ext}$ and
    $r_+=r_{1ext}$ which are the real roots of $T=0$. There is a point
    of type-2 phase transition located at $r_+=r_0$, at which the black
    hole heat capacity diverges. The physical black holes with positive
    temperature are unstable (Fig. 5). (\emph{III}) As it is shown in
    Fig. 6, the black holes with $\beta=1$ undergo type-2 phase
    transition at the divergent point of black hole heat capacity
    labeled by $r_+= r_1$. There is point of type-1 phase transition
    located at $r_+=r_{ext}$ where the black hole heat capacity
    vanishes. This class on new black hole solutions are stable provided
    that their horizon radii be in the range $r_{ext}<r_+<r_{1}$.

    Finally, making use of the grand canonical ensemble and noting the
    Gibbs free energy of the black holes, we analyzed the global
    stability of the new black holes obtained here. We determined the
    points at which the black holes experience Hawking-Page phase
    transition . Also we showed that there are some specific intervals
    for the horizon radii in such a way that the black holes with the
    horizon radius in this intervals are globally stable (Figs. 7-9).\\

    \textbf{Acknowledgement}: The authors appreciate the Research
    Council of Razi University for official support of this work.

    \setcounter{equation}{0}
    \begin{appendix}

        \setcounter{equation}{0}
        \section{Detailed derivation of Eq.(III.13)}

        With the purpose of finding the black hole mass for the spherically
        symmetric black holes, with the non-flat and non-AdS asymptotic
        behavior, identified in this work we start by substituting the
        metric function $W(r)$ into Eqs.(III.11) and (III.12). To do this, we
        consider the cases corresponding to the $\beta\neq1,\;\sqrt{3}$,
        $\beta=\sqrt{3}$, and $\beta=1$, separately.

        \subsection{The case $\beta\neq1,\;\sqrt{3}$}

        In this case, the quasilocal black hole mass can be obtained as
        follows \b
        {\cal{M}}=r\left(\frac{b}{r}\right)^{2\beta\gamma}(1-\beta\gamma)\left[\left\{\left[-mr^{2\beta\gamma-1}+u(r)\right]u(r)\right\}^{1/2}+mr^{2\beta\gamma-1}-u(r)\right],\e
        in which
        \b
        u(r)=\frac{1+\beta^2}{1-\beta^2}\left(\frac{r}{b}\right)^{2\beta\gamma}-\frac{\Lambda
            b^{2}(1+\beta^2)^2}{3-\beta^2}\left(\frac{r}{b}\right)^{\frac{2\gamma}{\beta}}
        +\frac{q^{2p}2^{p-1}(1+\beta^2)\Upsilon(\beta)}{(B-1)b^{^{2(^pB-1)}}}\left(\frac{b}{r}\right)^{2\eta}.\e
        Now, Eq.(A.1) can be rewritten as $$
        {\cal{M}}=(1-\beta\gamma)b^{2\beta\gamma}r^{1-2\beta\gamma}\left[u(r)\left\{1-\frac{mr^{2\beta\gamma-1}}{u(r)}\right\}^{1/2}+mr^{2\beta\gamma-1}-u(r)\right],$$
        $$=\frac{b^{2\beta\gamma}}{1+\beta^2}r^{1-2\beta\gamma}\left[u(r)\left\{1-\frac{mr^{2\beta\gamma-1}}{2u(r)}-\frac{1}{8}\left(\frac{mr^{2\beta\gamma-1}}{u(r)}\right)^2
        +{\cal{O}}\left(\frac{r^{2\beta\gamma-1}}{u(r)}\right)^3\right\}+mr^{2\beta\gamma-1}-u(r)\right],$$

        $$=\frac{b^{2\beta\gamma}}{1+\beta^2}r^{1-2\beta\gamma}\left[u(r)-\frac{mr^{2\beta\gamma-1}}{2}-\frac{u(r)}{8}\left(\frac{mr^{2\beta\gamma-1}}{u(r)}\right)^2
        +r^{2\beta\gamma-1}{\cal{O}}\left(\frac{r^{2\beta\gamma-1}}{u(r)}\right)^2+mr^{2\beta\gamma-1}-u(r)\right],$$
        \b=\frac{b^{2\beta\gamma}}{1+\beta^2}\left[\frac{m}{2}-\frac{m^2}{8}\left(\frac{r^{2\beta\gamma-1}}{u(r)}\right)
        +{\cal{O}}\left(\frac{r^{2\beta\gamma-1}}{u(r)}\right)^2\right].\e
        Thus, the black hole mass can be calculated as \b
        M=\lim_{r\rightarrow\infty}
        {\cal{M}}=\lim_{r\rightarrow\infty}\frac{b^{2\beta\gamma}}{1+\beta^2}\left[\frac{m}{2}-\frac{m^2}{8}\left(\frac{r^{2\beta\gamma-1}}{u(r)}\right)
        +{\cal{O}}\left(\frac{r^{2\beta\gamma-1}}{u(r)}\right)^2\right].\e
        It is easily shown that
        $\lim_{r\rightarrow\infty}\left(\frac{r^{2\beta\gamma-1}}{u(r)}\right)$
        and its higher powers are equal to zero. As the result one obtains
        \b M=\frac{b^{2\beta\gamma}}{1+\beta^2}\frac{m}{2}.\e

        \subsection{The case $\beta=\sqrt{3}$}

        In this case, regarding Eqs.(II.21) and (III.11), the black hole
        quasilocal mass is obtained as \b
        {\cal{M}}=\frac{1}{4}b^{\frac{3}{2}}r^{-\frac{1}{2}}\left[\left\{\left[-mr^{\frac{1}{2}}+u_3(r)\right]u_3(r)\right\}^{1/2}+mr^{\frac{1}{2}}-u_3(r)\right],\e
        \b
        u_3(r)=-2\left(\frac{r}{b}\right)^{2/3}-4\Lambda(b^3r)^{\frac{1}{2}}
        \ln\left(\frac{r}{L}\right)+\frac{2^{^{p+1}}q^{2p}\Upsilon(\beta=\sqrt{3})}{({\cal{B}}-1)b^{^{2(^p{\cal{B}}-1)}}}\left(\frac{b}{r}\right)^{2\xi}.\e
        Making use of the binomial expansion relation, after some algebraic
        simplifications, we arrive at
        \b{\cal{M}}=\frac{b^{\frac{3}{2}}}{8}-\frac{m^2b^{\frac{3}{2}}}{32}\left(\frac{r^{\frac{1}{2}}}{u_3(r)}\right)+{\cal{O}}\left(\frac{r^{\frac{1}{2}}}{u_3(r)}\right)^2.\e
        Taking limit $r\rightarrow\infty$, results in \b
        M=\frac{mb^{\frac{3}{2}}}{8}.\e

        \subsection{The case $\beta=1$}

        Noting Eq.(III.11) the quasilocal mass of the black holes, with the
        metric function given by Eq.(II.21), can be written in the following
        form \b
        {\cal{M}}=\frac{b}{2}\left[u_1(r)\left\{1-\frac{m}{u_1(r)}\right\}^{1/2}+m-u_1(r)\right],\e
        with \b
        u_1(r)=2\left[2-b^2(\Lambda+\lambda_1)+\ln\left(\frac{b}{r}\right)\right]\left(\frac{r}{b}\right)
        +\frac{p\;2^{^{p+1}}q^{2p}}{B_1(B_1-1)b^{^{2(^pB_1-1)}}}\left(\frac{b}{r}\right)^{B_1-1}.\e
        After some algebraic simplifications the quasilocal mass ${\cal{M}}$
        can be written as \b {\cal{M}}=\frac{m b}{4}+\frac{m^2
            b}{8}\frac{1}{u_1(r)}+{\cal{O}}\left(\frac{1}{u_1(r)}\right)^2,\e
        and by taking limit $r\rightarrow\infty$, we obtain \b M=\frac{m
            b}{4}.\e By summarizing Eqs.(A.5), (A.9) and (A.13), the analogous
        ADM black hole mass can be written in the general form given by
        Eq.(III.13).

    \end{appendix}

\end{document}